\documentclass[twocolumn,showpacs,secnumarabic,tightenlines,amssymb,
amsmath,nobibnotes,aps,prb]{revtex4}
\usepackage{graphicx}% Include figure files
\usepackage{dcolumn}% Align table columns on decimal point
\usepackage{bm}% bold math
\begin{document}
\preprint{APS/123-QED}
\title{Perturbative Approach to the Nonequilibrium Kondo Effect in a Quantum Dot}
% Force line breaks with \\
\author{Tatsuya Fujii$^1$}
%\altaffiliation[Also at ]{}
\author{Kazuo Ueda$^{1,2}$}%
%\email{}
\affiliation{$^1$Institute for Solid State Physics, 
University of Tokyo, Kashiwa 277-8581, Japan \\
$^2$Advanced Science Research Center, Japan Atomic Energy 
Research Institute, Tokai 319-1195, Japan
}
%\affiliation{$^2$Advanced Science Research Center, Japan Atomic Energy 
%Research Institute, Tokai, Ibaraki 319-1195, Japan
%}
%\author{}
%\homepage{}
%\affiliation{
\date{\today}
\begin{abstract}
The theory of quantum transport through a dot under a finite bias
voltage is developed using perturbation theory in 
the Keldysh formalism.  It is found that the Kondo resonance
splits into double peaks
when the voltage exceeds the Kondo temperature, $eV>k_B T_K $, which leads
to the appearance of a second peak in conductance, in addition to 
the zero-bias peak. The possible 
relevance of the new peak to the 0.7 conductance anomaly observed 
in quantum point contact is discussed. 
\end{abstract}
\maketitle
\section{INTRODUCTION}
Ten years after the theoretical predictions of the Kondo 
effect in a quantum dot\cite{Ng,Glazman}, the zero-bias peak of the 
differential conductance was identified as the Kondo resonance
\cite{Goldhaber,Cronenwett}. 
The observation that the peak value reached the 
unitary limit of $2e^2/h$ with decreasing temperature established
unambiguously that the Kondo effect is relevant to the transport
properties of the quantum dot\cite{Wiel}. 
Clearly the new feature 
of Kondo transport compared with the usual Kondo effect of 
magnetic impurities is the nonequilibrium nature, since the current
is measured with a finite bias voltage.  Only in the limit of 
zero bias voltage is the equilibrium condition recovered. 

This theoretical study of the nonequilibrium Kondo effect is based on 
the Keldysh formalism\cite{Keldysh}. 
In order to treat the correlation 
effect various methods have been used: one is perturbation theory
with respect to the Coulomb interaction $U$ in the dot\cite{Hershfield} 
and another often used method is the non-crossing approximation (NCA) 
where infinite 
$U$ is assumed\cite{Meir}. 
Concerning the equilibrium Kondo problem, it is 
well known that second order perturbation theory gives 
remarkably good results\cite{YandY}.  For example, the density of states
with the Kondo resonance in the middle of the side peaks corresponding 
to the energy levels in the atomic limit is well described by 
the theory.  However, it is not clear whether second order 
perturbation theory works well in nonequilibrium conditions. 
On the other hand, concerning the NCA, it is well known that 
the analyticity is broken in the low temperature limit.  Thus 
one cannot discuss the conductance in the unitary limit 
using the NCA. 

One important theoretical issue is that 
the above mentioned theories predict contradictory results
for the density of states (DOS) in the nonequilibrium case. The
NCA predicts splitting of the Kondo resonance under a finite 
bias voltage. 
On the other hand, in second order perturbation 
theory, the Kondo resonance peak is 
simply suppressed and does not show any particular structure 
in the nonequilibrium situation.  Since the Kondo resonance is 
a manifestation of singlet formation between the localized state
and the leads, it seems that the double peak structure at 
chemical potentials of both leads is reasonable. 
Indeed, the double peaks are also obtained by 
other approaches: equations of motion~\cite{Meir},
real-time diagrammatic formulation\cite{Koenig} and 
scaling methods\cite{Rosch2}. 
Moreover, it is not 
clear how the effects of the double peak structure, if it is present, 
would appear in the differential conductance.  
Recently, a proposal was made 
to measure the splitting of the DOS by using three-terminal quantum 
dots\cite{Sun,Lebanon}. 
As a matter of fact, the splitting of the Kondo peak 
was successfully observed by introducing a potential difference in the
source lead\cite{Fran}. 
This result is most likely due to the splitting of the 
Kondo resonance, although the precise geometry is different from 
the present case where a finite voltage is applied 
between the source and drain. 
It is clearly necessary to 
study nonequilibrium Kondo effects by a better theoretical 
approach.

In this paper we analyze the Kondo effect  
in a quantum dot with a finite voltage 
by using perturbation theory up to fourth order in $U$ 
based on the Keldysh formalism.
We show that with increasing bias voltage 
a single Kondo peak splits into double peaks 
at $eV \sim k_B T_K$. 
As a result, 
an anomalous peak of the differential conductance 
appears when $eV>k_B T_K $. 
At the end of this paper, we will also 
discuss the possible relevance to 
the experiments referred to 
as the $0.7$ conductance anomaly 
in quantum point contacts (QPC)\cite{Thomas,Kris,Crone2}.
\section{MODEL and CALCULATIONS}

We consider a single-level quantum dot attached to 
two leads. 
This system is described by the Anderson impurity model, 
%%%%%%%%%%%%%%%%%%%%%%%%%%%%%%%%%%%%%%%%%%%%%%%%%%%%%%%
\begin{eqnarray}
 H= \sum_{k\alpha\sigma} 
     {\varepsilon}_{k \alpha} 
      c^{\dagger}_{k\alpha\sigma} c_{k\alpha\sigma}
      +\sum_{\sigma} {\epsilon}_d n_{\sigma}
       +U n_{\uparrow} n_{\downarrow} & &
             \nonumber\\
      + \sum_{k\alpha\sigma}
          \left(V_{k\alpha\sigma} c^{\dagger}_{k\alpha\sigma} d_{\sigma}
           +{\rm h.c.}\right),& &
\label{Hamiltonian}
\end{eqnarray}
%%%%%%%%%%%%%%%%%%%%%%%%%%%%%%%%%%%%%%%%%%%%%%%%%%%%%%%%
where $\alpha=L,R$ and $c_{kL\sigma}$ ($c_{kR\sigma}$) annihilates 
an electron in the left(right)-lead,
$d_{\sigma}$ annihilates an electron with spin $\sigma$ in the dot, 
and
$n_{\sigma}=d^\dagger_{\sigma}d_{\sigma}$.
The coupling constants $V_{k\alpha\sigma}$ describe the tunneling 
matrix elements between the dot and leads. 
Nonequilibrium situation is driven by a finite difference between 
$\mu_L$ and $\mu_R$ which are the chemical potentials of the leads in 
both sides. 
%%The difference of $\mu_L -\mu_R$ 
%%corresponds to the potential drop, $eV$ at the dot. 

For simplicity 
we concentrate on 
the symmetric Anderson model, 
where the dot is symmetrically coupled with the leads and 
the energy level of the dot including 
the Hartree mean field $U\bar{n}_\sigma$ coincides with 
the center of the potential drop, $eV=\mu_L -\mu_R$. 
%%When we set the origin of energy at the middle of the potential drop, 
Then the symmetric conditions are stated as 
$\Gamma_{L\sigma}=\Gamma_{R\sigma} \equiv \Gamma$, 
$\mu_L= -\mu_R =eV/2$ and $-{\epsilon}_d={\epsilon}_d +U$. 
Here $\Gamma_{L,R\sigma}$ represent 
resonance width at the chemical potentials, 
$\Gamma_{L,R\sigma}(\omega)=2\pi\sum_k |V_{kL,R\sigma}|^2 \delta 
(\omega -\varepsilon_{k L,R})$. 
In this paper we restrict ourselves to the ground state, $T=0$. 

Our first aim is to calculate the density of states in the dot, 
\begin{eqnarray}
\rho_{\sigma}(\omega)=-\frac{1}{\pi}{\rm Im}G^r_{\sigma} (\omega), 
\end{eqnarray}
where $G^r_{\sigma} (\omega)$ is the retarded Green function. 
Then the current through the dot 
is expressed by 
\begin{eqnarray}
I=\frac{e}{\hbar} \sum_{\sigma} \int^{\infty}_{-\infty} d\omega
%\bar{\Gamma_{\sigma}}(\omega)
\frac{\Gamma_{L\sigma} \Gamma_{R\sigma}}
{\Gamma_{L\sigma} +\Gamma_{R\sigma}}
\rho_{\sigma}(\omega)
(f_L(\omega)-f_R(\omega)),
\end{eqnarray}
where $f_{L,R}(\omega)=\theta (\mu_{L,R}-\omega)$. 
%%1/(1+e^{\beta (\omega -\mu_{L,R})})$.
The differential conductance is obtaied 
from the current by 
\begin{eqnarray}
G(V)=\frac{\partial I}{\partial V}. 
\label{gv}
\end{eqnarray}
\section{KELDYSH FORMALISM}
\subsection{PERTURBATION THEORY}

Now we briefly sketch perturbation theory 
based on the Keldysh formalism\cite{Keldysh}. 
Each of terms generated by the expansion of the S matrix 
includes integrals along the Keldysh contour 
which starts at $t=-\infty$, passes through $t=\infty$ and 
returns back to $t=-\infty$. 
The branch from $t=-\infty (t=\infty)$ to 
$t=\infty (t=-\infty)$ is denoted by the index $-(+)$. 
Thus we need to introduce four types of 
Green functions with these additional indices 
$G^{\alpha \beta}$ where $\alpha$ and $\beta$ are$-$ or $+$. 

The interaction terms in the S matrix are given by 
the sum of the Coulomb interaction $U$ at the dot and 
the hybridizations $V_{kL,R\sigma}$ between the leads and the dot. 
We employ perturbation theory in $U$ based on 
the Keldysh formalism 
where the Green functions in each order of $U$ are 
renormalized by $V_{kL,R\sigma}$. 

The Green functions of $0$th order in $U$ can be explicitly 
evaluated as 
\begin{eqnarray}
%%& &g^{0-+}_{\sigma}(\omega) =- i 2\pi |g^{0r}_{\sigma}(\omega)|^2
%%\sum_{k p=L,R} |V_{kp\sigma}|^2 \delta (\omega -\varepsilon_k) 
%%f_p (\varepsilon_k),\\
%%& &g^{0+-}_{\sigma}(\omega) =+ i 2\pi |g^{0r}_{\sigma}(\omega)|^2
%%\sum_{k p=L,R} |V_{kp\sigma}|^2 \delta (\omega -\varepsilon_k) 
%%(1-f_p (\varepsilon_k)),
%%
& &g^{0-+}_{\sigma}(\omega) =- i |g^{0r}_{\sigma}(\omega)|^2
\sum_{p=L,R} \Gamma_{p\sigma}(\omega)
f_p (\omega),\\
& &g^{0+-}_{\sigma}(\omega) =+ i |g^{0r}_{\sigma}(\omega)|^2
\sum_{p=L,R} \Gamma_{p\sigma}(\omega)
(1-f_p (\omega)),
\end{eqnarray}
by solving the Dyson equations concerning $V_{kL,R\sigma}$ 
where $g^{0r}_{\sigma}(\omega)=(\omega -\epsilon_d +i/2 \cdot 
(\Gamma_{L\sigma}(\omega)+\Gamma_{R\sigma}(\omega))^{-1}$.
In order to estimate the diagonal components, 
it is more convenient to find 
the Fourier transform of 
\begin{eqnarray}
g^{0\alpha\alpha}_{\sigma}(t)=\theta (-\alpha t)g^{0+-}_{\sigma}(t)
+\theta (\alpha t)g^{0-+}_{\sigma}(t),
\end{eqnarray}
rather than 
to solve $g^{0\alpha\alpha}_{\sigma}$ directly. 

In the expansion in $U$ there are diagrams 
in which Hartree type self-energies are inserted. 
Their contributions are entirely taken into account by 
substituting 
$\epsilon_d \rightarrow \epsilon_d +U\bar{n}_{-\sigma}$ in 
$g^{0\alpha \beta}_{\sigma}$ in the diagrams 
where the Hartree type of self-energies are omitted. 
Using $\epsilon_d +U\bar{n}_{-\sigma}=0$ for the symmetric case, 
the new Green functions $g^{\alpha \beta}_{\sigma}$ 
are 
\begin{eqnarray}
& &g^{-+}_{\sigma}(\omega) =- i 2{\rm Im}g^{r}_{\sigma}(\omega)
f_{\rm eff}(\omega), \\
& &g^{+-}_{\sigma}(\omega) =+ i 2{\rm Im}g^{r}_{\sigma}(\omega)
(1-f_{\rm eff}(\omega)), \\
& &g^{--}_{\sigma}(\omega)=\frac{1-f_{\rm eff}(\omega)}{\omega + i \Gamma}+
\frac{f_{\rm eff}(\omega)}{\omega - i \Gamma}, \\
& &g^{++}_{\sigma}(\omega)=-g^{--}_{\sigma}(\omega)^* ,
\end{eqnarray}
where 
$\Gamma_{L,R\sigma}(\omega)$ 
approximated the values at the chemical potentials, 
$f_{\rm eff}(\omega)=
(\Gamma_{L\sigma} f_L (\omega)+\Gamma_{R\sigma} f_R (\omega))/
(\Gamma_{L\sigma} +\Gamma_{R\sigma})
$ and 
$g^{r}_{\sigma}(\omega)=(\omega + i \Gamma)^{-1}$. 

Then the Dyson equations for $U$ may be written in matrix form, 
\begin{eqnarray}
G^{\alpha \beta}_{\sigma}(\omega)=g^{\alpha \beta}_{\sigma}(\omega)
+g^{\alpha \gamma}_{\sigma}(\omega)
\Sigma^{\gamma \delta}_{\sigma}(\omega)
G^{\delta \beta}_{\sigma}(\omega).
\label{dyson}
\end{eqnarray}
Here in each order of $\Sigma^{\gamma_1 \gamma_2}_{\sigma}$ 
it is sufficient to exclude Hartree type self-energies. 

Let us evaluate the self-energies by using perturbation theory 
up to the fourth order of $U$. 
It is known that the perturbative expansion for 
the Anderson model\cite{YandY} is effective and well behaved 
in the equilibrium case, 
since 
the exact solution shows a rapid convergence of the perturbation series
\cite{Zlatic}. 
However, in the nonequilibrium situation, 
analysis of perturbation theory 
has been limited to second lowest order. 
To proceed to higher order calculations, 
the four-point vertex, 
which is obtained by extending 
the method used by Keldysh for the electron-phonon vertex\cite{Keldysh}, 
is very convenient. 
This procedure is to insert vertices , 
\begin{eqnarray}
\Gamma^{(0)\alpha_1 \alpha_2,\alpha_3 \alpha_4}=
U(\gamma_{\alpha_1 \alpha_2}^{-} \gamma_{\alpha_3 \alpha_4}^{-}-
  \gamma_{\alpha_1 \alpha_2}^{+} \gamma_{\alpha_3 \alpha_4}^{+} ), 
\end{eqnarray}
in each of diagrams, 
which may be called {\it the Keldysh vertices}. 
Here
$
\gamma_{\alpha_1 \alpha_2}^{\alpha_3}
=\delta_{\alpha_1 \alpha_2} \sigma^{z}_{\alpha_2 \alpha_3},
$
where 
$\sigma^{z}_{\alpha_2 \alpha_3}$ is the third Pauli matrix. 

\begin{figure}[h]
\includegraphics[width=7cm]{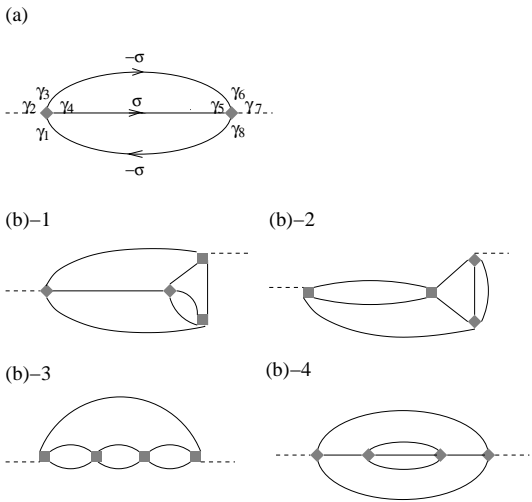}
\caption{\label{1} (a) the second-order diagram and (b) the 
fourth-order diagrams. 
In (b) all twelve diagrams are generated by including 
the directions of particle lines and spins. 
Shaded squares represent the Keldysh vertices.}
\label{diagram}
\end{figure}

All diagrams up to fourth order are shown in Fig.~\ref{diagram} 
for the symmetric case, where the third-order terms 
vanish in the same way as 
the equilibrium case. 
%%%%%%%%%%%%%%%%%%%%%%%%%%%%%%%%%%%%%%%%%%%%%%%%%%%%%%%%%%%%
%
%%%%%%%%%%%%%%%%%%%%%%%%%%%%%%%%%%%%%%%%%%%%%%%%%%%%%%%%%%%%
Fig.~\ref{diagram}(a) is the second-order diagram 
in which directions of particle lines, spins and 
the Keldysh vertices are depicted. 
In this case, the second order correction of the self-energies 
is 
\begin{eqnarray}
& &\Sigma^{(2)\gamma_7 \gamma_2}_{\sigma}(\omega)=
i^2 (-1) \int^{\infty}_{-\infty}
\frac{d\omega_1}{2\pi} \frac{d\omega_2}{2\pi}
\Gamma^{(0)\gamma_1 \gamma_2,\gamma_3 \gamma_4} \\
& &g^{\gamma_5 \gamma_4}_{\sigma}(\omega_1) 
g^{\gamma_6 \gamma_3}_{-\sigma}(\omega_2)
g^{\gamma_1 \gamma_8}_{-\sigma}(\omega_1+\omega_2 -\omega)
\Gamma^{(0)\gamma_5 \gamma_6,\gamma_7 \gamma_8}.
\nonumber
\end{eqnarray}

We can proceed in the same way 
to the fourth-order diagrams shown in Fig.~\ref{diagram}(b). 
Four different representative diagrams are 
illustrated where directions and spin variable are not specified. 
By specifying them, 
one can immediately obtain all twelve diagrams.
Note that Fig.~\ref{diagram}(b)-1,2,3 and Fig.~\ref{diagram}(b)-4 
are skeleton and non-skeleton 
diagrams, respectively. 
%%Here we will write down one of three diagrams 
%%which are derived by drawing directions and spin 
%%in Fig.~1(b)-1. 
%%\begin{widetext}
%%\begin{eqnarray}
%%\nonumber
%%& &\Sigma^{(4a)\gamma_7 \gamma_2}_{\sigma}(\omega)=
%%i^4 (-1)^2\int^{\infty}_{-\infty}
%%\frac{d\omega_1}{2\pi} \frac{d\omega_2}{2\pi}
%%\Gamma^{(0)\gamma_1 \gamma_2,\gamma_3 \gamma_4}
%%g^{\gamma_5 \gamma_4}(\omega_1)g^{\gamma_6 \gamma_3}(\omega_2)
%%g^{\gamma_1 \gamma_8}(\omega_1+\omega_2 -\omega) \\
%%%
%%& &\int^{\infty}_{-\infty}
%%\frac{d\omega_3}{2\pi} \frac{d\omega_4}{2\pi}
%%\Gamma^{(0)\beta_1 \gamma_5,\beta_2 \gamma_7}
%%g^{\delta_1 \beta_2}(\omega_1+\omega_3 -\omega_{2})
%%g^{\beta_1 \alpha_1}(\omega_3)
%%%
%%\Gamma^{(0)\delta_2 \delta_1,\gamma_8 \delta_3}
%%g^{\delta_2 \alpha_2}(\omega_2-\omega_3 +\omega_4)
%%g^{\alpha_3 \delta_3}(\omega_4)
%%\Gamma^{(0)\alpha_3 \gamma_6,\alpha_1 \alpha_2}.
%%\end{eqnarray}
%%\end{widetext}
%
It is tedious but straightforward to evaluate 
the fourth-order contributions, 
$
\Sigma^{(4)\gamma_1 \gamma_2}_{\sigma}
$, 
in the same manner as the second order. 

From the self-energy matrix , 
the imaginary part of $\Sigma^r_{\sigma}$ is defined by,
\begin{eqnarray}
{\rm Im}\Sigma^r_{\sigma} (\omega)=
\frac{1}{2} \left( i \Sigma^{+-}_{\sigma}(\omega)
- i \Sigma^{-+}_{\sigma}(\omega) \right).
\end{eqnarray}
Then the real part of $\Sigma^r_{\sigma}$ 
is obtained by the Kramers-Kronig relation, 
\begin{eqnarray}
{\rm Re}\Sigma^r_{\sigma} (\omega)=
\frac{1}{\pi}{\cal P}\int^{\infty}_{-\infty}
\frac{{\rm Im}\Sigma^r_{\sigma}
 (\omega^{'})}{\omega^{'} -\omega} d\omega^{'}.
\label{Kramers}
\end{eqnarray}
The Dyson equation for the retarded component 
is derived from eq.(\ref{dyson}) as 
\begin{eqnarray}
G^{r}_{\sigma}(\omega)=g^{r}_{\sigma}(\omega)
+g^{r}_{\sigma}(\omega)
\Sigma^{r}_{\sigma}(\omega)G^{r}_{\sigma}(\omega).
\end{eqnarray} 
\subsection{SELF-ENERGIES}

\begin{figure}[h]
\includegraphics{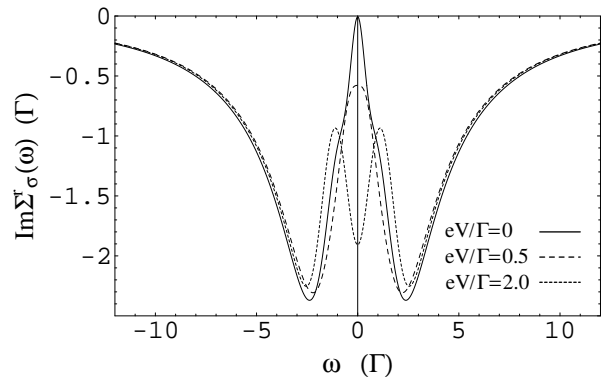}
\caption{The numerically calculated 
${\rm Im}\Sigma^r_{\sigma}$ 
as a function of $\omega$ for $eV/\Gamma=0$, 
$0.5$ and $2.0$. In this example $U/\Gamma$ 
is $6$.}
\label{ims}
\end{figure}

In Fig.~\ref{ims} the numerically calculated 
$
{\rm Im}\Sigma^r_{\sigma} (\omega)
$
is shown for $U/\Gamma=6$. 
%%%%%%%%%%%%%%%%%%%%%%%%%%%%%%%%%%%%%%%%%%%%%%%%%%%%%%%%%%%%
%
%%%%%%%%%%%%%%%%%%%%%%%%%%%%%%%%%%%%%%%%%%%%%%%%%%%%%%%%%%%%
In the equilibrium case of $eV/\Gamma=0$, 
${\rm Im}\Sigma^r_{\sigma}|_{eV/\Gamma=0}$ 
is dominated by the second-order contribution 
as is well known\cite{YandY}, 
since the fourth-order contributions of skeleton diagrams shown 
in Fig.~\ref{diagram}(b)-1,2,3 and 
nonskeleton diagram in Fig.~\ref{diagram}(b)-4 almost cancel 
each other except in the low energy limit. 
In order to analyze $\omega \sim 0$, 
it should first be noted that 
the present perturbation results correctly 
reproduce the low-energy asymptotic form in the equilibrium 
limit\cite{YandY}: 
\begin{eqnarray}
\nonumber
& & {\rm Im}\Sigma^r_{\sigma} (\omega) |_{eV/\Gamma =0} \\
& & {\quad \quad} \simeq 
\frac{\Gamma}{2} \left\{ -\left( \frac{U}{\pi\Gamma} \right)^2 
-3( 10- \pi^2 ) \left( \frac{U}{\pi\Gamma} \right)^4 \right\}
\left( \frac{\omega}{\Gamma} \right)^2.
\nonumber
\end{eqnarray}
The nonskeleton diagram does not give any contribution 
to the fourth-order term of the coefficient of $(\omega /\Gamma)^2$. 
Among the skeleton diagrams of the RPA-type diagrams, 
Fig.~\ref{diagram}(b)-3, give $(-3\times 3)$ for the coefficient 
of $(U/\pi \Gamma)^4$, while those from the vertex-correction-type, 
Fig.~\ref{diagram}(b)-1 and 2, give 
$3\times (\pi^2-7)$, resulting in the small number of 
$-3\times (10-\pi^2)$. 
The way that cancellation occurs between the RPA-type diagrams and 
the vertex-correction-type diagrams changes under a finite bias voltage. 

%%%%%%%%%%%%%%%%%%%%%%%%%%%%%%%%%%%%%%%%%%%%%%%%%%%%%%%%%%%%
\begin{figure}[h]
\includegraphics{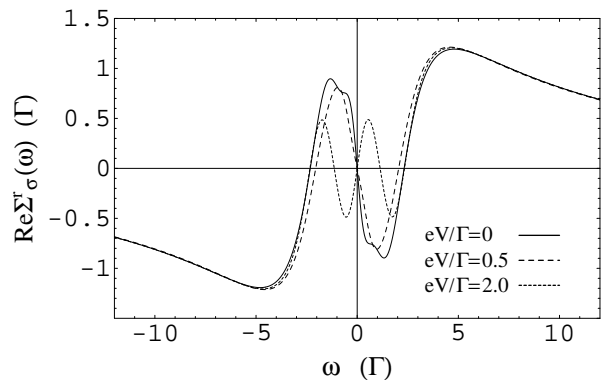}
\caption{
The real part of the self-energy 
for $eV/\Gamma=0$, $0.5$ and $2.0$ 
with $U/\Gamma =6$.}
\label{res}
\end{figure}
%
%%%%%%%%%%%%%%%%%%%%%%%%%%%%%%%%%%%%%%%%%%%%%%%%%%%%%%%%
In the vicinity of the equilibrium (for example at $eV/\Gamma=0.5$), 
$-{\rm Im}\Sigma^r_{\sigma} (\omega =0)$ becomes 
slightly larger  
but still keeps the structure of the equilibrium case, 
which means that damping of quasi-particles 
is simply enhanced by real transitions between the leads and the dot. 
Novel features are observed 
with further increasing the voltage. 
The fourth-order contribution to 
the coefficient of the $(\omega /\Gamma)^2$ term 
changes sign. 
Even though the second-order contribution remains negative, 
the coefficient of $(\omega /\Gamma)^2$ becomes positive for 
sufficiently large $U/\pi \Gamma$. 
Therefore, 
the peak of ${\rm Im}\Sigma^r_{\sigma}$ around $\omega \sim 0$ 
becomes depressed.
In this case the quasi-particles remain coherent 
around $\omega=\pm eV/2$ as is shown for $eV /\Gamma =2.0$

Now we turn to ${\rm Re}\Sigma^r_{\sigma}$ 
which is shown in Fig.~\ref{res}. 
In the case of $eV/\Gamma=2.0$ we see 
the development of new zero-points of 
${\rm Re}\Sigma^r_{\sigma}$ at $\omega =\pm eV/2$
which are absent for $eV/\Gamma=0$ and $0.5$. 
Since the slopes at the zero-points are negative, 
which leads to mass enhancement, the effect of damping 
due to ${\rm Im}\Sigma^r_{\sigma}$, is reduced. 
\subsection{DENSITY OF STATES}

\begin{figure}[h]
\includegraphics{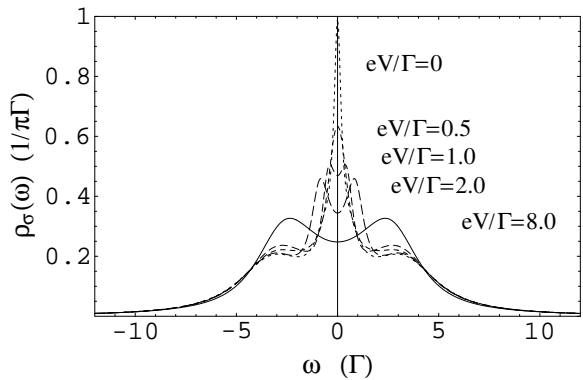}
\caption{\label{3} The density of states in units of 
$1/\pi\Gamma$ for various $eV/\Gamma$ 
with $U/\Gamma=6$.}
\label{dos}
\end{figure}

The density of states is shown 
in Fig.~\ref{dos} in units of 
$1/\pi\Gamma$. 
In the equilibrium case, 
the sharp Kondo peak develops in the middle of 
the two peaks at around $\pm U/2$. 
With a finite voltage, 
the Kondo peak 
at $\omega /\Gamma=0$ is suppressed 
but still keeps a {\it single} peak with broadening 
for small $eV/\Gamma$, in accordance with the 
analysis based on the Ward identities\cite{Oguri}. 
When the voltage is further increased, 
the Kondo peak splits into {\it double} peaks at $\omega \sim \pm eV/2$, 
which are located near 
the chemical potentials of the two leads. 
This behavior is qualitatively consistent with the results 
obtained by the NCA\cite{Meir}. 
This change occurs when the potential drop $eV$ exceeds 
$k_B T_K$, defined as the full-width at half-maximum of the Kondo peak. 
For $U/\Gamma =6$, $k_B T_K /\Gamma$ is estimated to be about $0.6$. 
%%%%%%%%%%%%%%%%%%%%%%%%%%%%%%%%%%%%%%%%%%%%%%%%%%%%%%%%%%%%
%
%%%%%%%%%%%%%%%%%%%%%%%%%%%%%%%%%%%%%%%%%%%%%%%%%%%%%%%%%%%%

With higher bias the {\it double} 
Kondo peaks located at $\omega \sim \pm eV/2$ 
merge with the peaks at the atomic limit 
$\omega \sim \pm U/2$ and then the latter 
dominate when $eV > U$. 
We have checked that this structure coincides with 
the result obtained by the second order perturbation theory. 
The origin of this phenomena is that 
the higher-order scatterings become unimportant in the high voltage 
regime due to the strong dissipative processes 
between the leads and the dot. 
This result is consistent with ref.\onlinecite{Rosch2} in that 
the nonequilibrium decoherence destroys the Kondo effect 
when $eV >> k_B T_K$. 
\section{CONDUCTANCE}

\begin{figure}[h]
\begin{center}
\includegraphics{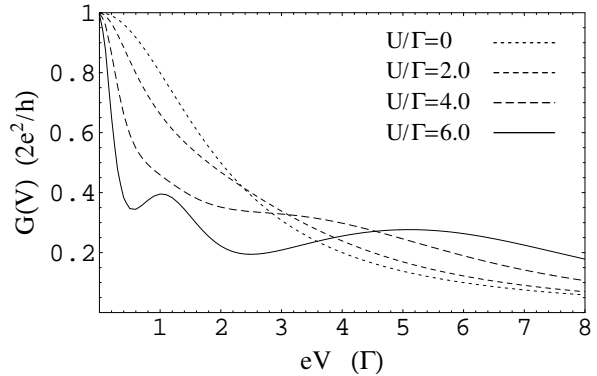}
\caption{\label{4} The differential conductance in units of $2e^2/h$
for various values of $U/\Gamma$. 
The zero-bias peaks reach the unitary limit $2e^2/h$. 
For $U/\Gamma =6$, the anomalous peak is developed 
in $eV>k_B T_K $ 
in addition to the zero bias peak. }
\label{dfc}
\end{center}
\end{figure}

In Fig.~\ref{dfc} the differential conductance 
defined by eq.(\ref{gv}) 
is shown in units of $2e^2/h$ for various values of 
$U/\Gamma$ as a function of bias voltage.  
For all $U$, the zero-bias peak starts from the unitary
limit.  As $U$ is increased, 
the width of the zero-bias peak becomes narrower. 
For $U/\Gamma= 4$ and $6$, 
a broad peak is seen at around $eV \sim U$. 
This broad peak corresponds
to tunneling processes through the energy levels 
in the atomic limit.  Between the zero-bias peak and the broad peak, 
a new peak appears for large $U$ ($U/\Gamma = 6$ in the figure).  
%%%%%%%%%%%%%%%%%%%
%
%%%%%%%%%%%%%%%%%%%%%%%%%%%%%%%%%%%%%%%%%%%%%%%%%%%%%%%%%%%%

To understand the origin of the new peak, first we
rewrite the expression of the conductance, eq.(\ref{gv}), as 
\begin{eqnarray}
\nonumber
G(V) &=&\frac{2e^2}{h}\pi \Gamma
\left( \rho_{\sigma} (eV/2)+\int^{eV/2}_{-eV/2}
\frac{\partial \rho_{\sigma} (\omega)}
{\partial eV} d \omega \right) \\ 
     &\equiv&G_1 (V)+G_2(V)
\end{eqnarray}
%
%%%%%%%%%%%%%%%%%%%%%%%%%%%%%%%%%%%%%%%%%%%%%%%%%%%%%%%%%%%%%%%%%%%
\begin{figure}[h]
\begin{center}
\includegraphics{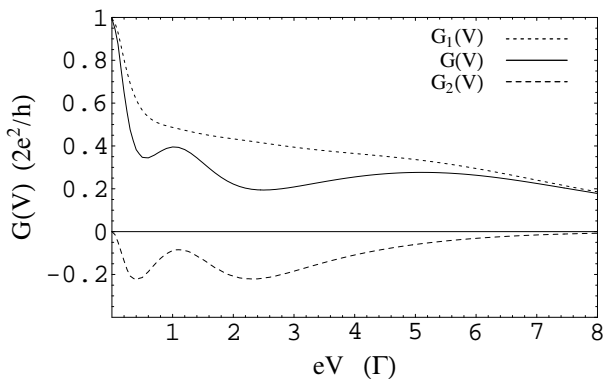}
\caption{\label{4} The differential conductance for $U/\Gamma =6$. 
$G(V)$, is given by the sum of $G_1 (V)$ and $G_2 (V)$. 
The new peak originates in the behavior of $G_2 (V)$.
}
\label{np}
\end{center}
\end{figure}
%
%%%%%%%%%%%%%%%%%%%%%%%%%%%%%%%%%%%%%%%%%%%%%%%%%%%%%%%%%%%%%%%%%%%
As shown in Fig.~\ref{np}, 
the first term, $G_1 (V)$, decreases monotonically as $V$ is 
increased. Concerning the second term $G_2 (V)$ first we note that
if the range of integration 
integrand is the entire frequency space, 
the integral should vanish
due to the sum-rule of the spectral weight.  For $eV < k_BT_K$ 
the spectral weight shifts 
to higher frequencies, giving a negative 
contribution.  When $eV \sim k_BT_K$ a considerable part of positive 
$\partial \rho_{\sigma} (\omega)/\partial eV$
enters in the integrand, thus giving 
a less negative contribution. Therefore we may conclude 
that the new peak appears when the bias voltage exceeds 
the Kondo energy.
\section{DISCUSSIONS AND SUMMARY}

Finally we discuss the possible relevance of the present study 
to the $0.7$ conductance anomaly in QPC\cite{Thomas,Kris,Crone2}. 
It has been suggested that 
Kondo effect plays significant role 
as the origin of the $0.7$ structure in QPC 
because the structure is found 
when the unitary limit of $2e^2/h$\cite{Crone2} is approached 
with decreasing temperature. 
In view of the universal nature of the Kondo phenomena, it may be 
interesting to consider experimental results of the QPC 
in light of the present results. 
Fig. 3 of ref.\onlinecite{Kris} or Fig. 1 of ref.\onlinecite{Crone2} 
shows that at a gate voltage for 
which the zero-bias conductance reaches the unitary limit the 
conductance first drops with increasing the bias voltage and then 
starts to increase again, leading to a second peak at finite 
voltage. The envelope made of these second peaks for different  
gate voltages forms the 0.8 plateau. 
As the temperature is increased the 0.8 plateau 
is extrapolated to the 0.7 structure in the zero-bias limit. 
In fact, Fig. 3(b) of ref.\onlinecite{Crone2} clearly demonstrates 
that a second 
peak appears when $eV > k_BT_K$. 
This result seems to support the new peak obtained in 
the present study.

However the peak-height obtained by the present study 
does not reach to $0.8$ as is seen in Fig.~\ref{dfc}. 
This problem may be resolved in the future 
either by considering higher-order corrections 
or actual level schemes of the QPC. 

In summary we have studied the Kondo transport through a dot with 
a finite voltage by using perturbation theory in the 
Keldysh formalism. 
It is shown that the splitting of the Kondo resonance 
occurs when the bias voltage exceeds the Kondo temperature. 
As a result, the new peak in the differential conductance appears 
when the electron-electron correlation $U$ is sufficiently strong. 
Finally, we have suggested that the present results are relevant
also to the 0.7 conductance anomaly.  
Clearly further studies are necessary to elucidate the relation 
between the present analysis and the 0.7 anomaly in the QPC.
%In experiment of ref.
%\onlinecite{Crone2}, 
%it was clearly demonstrated that the crossover between the region of 
%the conductance of 
%the unitary limit and the 0.8 plateau region took place around 
%the Kondo temperature. This result was consistent with the prediction of 
%the present study. 
%
\section{ACKNOWLEDGEMENT}

The authors would like to thank K. Kobayashi and 
A. Oguri for helpful discussions. 
%%%%%%%%%%%%%%%%%%%%%%%%%%%%%%%%%%%%%%%%%%%%%%%%%%%%%%%%%%%%
%%%%%%%%%%%%%%%%%%%%%%%%%%%%%%%%%%%%%%%%%%%%%%%%%%%%%%%%%%%%

\end{document}